\documentclass[reprint,amsmath,amssymb,aps,prb]{revtex4-1}
\usepackage{epsfig,epstopdf}
\usepackage{bm,hyperref}
\hypersetup{pdfpagemode=UseNone}
\usepackage{lipsum}
\usepackage{graphicx}
\usepackage{amsfonts}
\usepackage{amssymb}
\usepackage{dcolumn}
\usepackage{amsmath}
\usepackage{soul}
\usepackage{color}
\usepackage{gensymb,textcomp}
\def\e{\begin{equation}}
\def\f{\end{equation}}
\def\-#1{{\bf #1}}
\def\o{\omega}
\def\va{\varepsilon}
\def\.{\cdot}

\def\=#1{\overline{\overline #1}}
\def\Re{{\rm Re\mit}}
\def\Im{{\rm Im\mit}}

\begin{document}

%%% Title of paper
\title{Dual-metasurface superlens: a comprehensive study}

\author{\underline{M.S.M. Mollaei}$^{1}$ and C. Simovski$^{1,2}$}

%%% Authors and affiliation - use for multiple authors
% (here given for 2 authors, please adjust to exact number and underline
% the corresponding author)
\affiliation{$^1$Department of Electronics and Nanoengineering, Aalto University, P.O.~Box 15500, FI-00076 Aalto, Finland\\
$^2$University ITMO, Kronverkski 47, 197101, St. Petersburg, Russia}

\date{\today}

\begin{abstract}
We present a theoretical {and numerical} study of a dual-metasurface superlens dedicated to the near-field optical imaging of submicron objects.
Compared to the previous studies of dual-metasurface plasmonic superlenses, we suggest a more adequate {theoretical model} of their
operation. The new model allows us to obtain twice better operational characteristics of the re-designed superlens. For the first time, we describe the operation
of such the superlens using full-wave numerical simulations, taking into account the interfaces of the host medium slab and proving the nanoimaging
for scattering objects instead of radiating sources. We discuss and address both application and fabrication issues for this superlens.
\end{abstract}

\maketitle

\section{Introduction}

In optical imaging the Abbe diffraction limit represents a problem when the target is to image the subwavelength details of a substantial object or to resolve two small objects \cite{Born}.
In the modern era of nanotechnology and nanoscience, significant efforts were done so that to beat this limit. In fact, if we eliminate optical noises the image can be restored using different types of postprocessing the recorded optical signal. Here a long list of techniques from time-reversal nanoimaging \cite{Fink} to quantum optical imaging \cite{Quant} could be presented. However,
in the present paper we concentrate on the imaging without post-processing achievable with a steady optical signal at a frequency within the the visible light spectrum.

In work \cite{Pen} a perfect imaging method has been suggested by Sir John Pendry which  implied the exact reproduction of the electromagnetic field distribution
over an object in the image area. For it, an object should be located in front of the so-called \emph{Veselago pseudolens} at the distance $\Delta$ equal to one half of its thickness.
The Veselago pseudolens is a flat layer of a hypothetic medium with permittivity $\va$ and permeability $\mu$
both equal to those of the host medium with sign minus. If the pseudolens is located in free space we have $\va=\mu=-1$.
Perfect imaging was predicted by Pendry even for the case when $\Delta/\lambda>1$. In accordance to the Pendry's theory, the Veselago pseudolens is
a \emph{perfect lens} -- a device which combines two functionalities. The first functionality is a reflection-and-aberration-free parallel-plate focusing of divergent wave beams,  predicted by Victor Veselago in \cite{Ves}. Note, that this functionality had been (prior to \cite{Pen}) implemented in a photonic-crystal layer whose response to propagating plane waves is the same as that of the Veselago pseudolens \cite{Sil}.  The second functionality of the perfect lens is the amplification of evanescent waves across the layer \cite{Pen}. Pendry has shown that
this amplification is uniform over the evanescent spatial spectrum that offers an exact compensation of the attenuation of evanescent waves over two paths -- from the source to the
perfect lens and from the last one to the image. Thus, the field in the image area exactly reproduces that in the object area {(see also in \cite{all,broadband,negative,near})}.

Unfortunately, to build the Veselago medium or at least to sufficiently approach to the regime $\va=\mu=-1$ in the optical range turned out to be an impossible mission.
Numerous attempts  failed due to the impact of the granularity of optical composites targeted to mimic the Veselago medium and especially due to the impact of optical losses corresponding to the regime ${\rm Re}(\mu)=-1$ in the optical range. However, work \cite{Pen} gave a tremendous pulse to the development of metamaterials -- composites with extreme and useful electromagnetic properties, not observed in natural media. In particular, Pendry gave birth to so-called \emph{superlenses}, also suggested in work \cite{Pen}. A superlens sacrifices the functionality of the far-field image
(correct reproduction of the object shape) in favor of the spatial resolution of the subwavelength details. The  superlens suggested in \cite{Pen} was a layer of material with
$\va=-1$ (for free space) or $\va=-\va_d$ if the epsilon-negative layer is sandwiched in the dielectric ambient of permittivity $\va_d$. Pendry suggested to use silver -- a material
with minimal $|\Im (\va)|$ at frequencies corresponding to $\Re (\va)=-\va_d$.

Further studies have shown that the issue of flatness is critical for this superlens -- only sub-nanometer corrugations are allowed. Development of nanotechnologies allowed
several scientific groups to demonstrate such the superlenses experimentally (see e.g. in \cite{Fang,Mel,experimentally}), and consequently
the topic of nanoimaging using the superlens has formed a body of literature. Initial optical superlenses such as \cite{Fang,Mel} demonstrated the spatial resolution $\delta$ of two subwabvelength objects close to $\lambda/6$ in the case when the distance $D$ between the centers of the object and its image areas was as small as $\lambda/5\dots \lambda/4$.
Practically, the distances $D$ in these negative-epsilon superlenses were within the interval $D=75-120$ nm that corresponded to the Ag layers of thicknesses $d=D-2\Delta=35-70$ nm. This result implied the sandwiched metal-dielectric structures (the subwavelength objects were embedded into the front dielectric layer and the images were obtained in the rear dielectric layer of photoresist). This approach is evidently unsuitable for imaging the moving objects.

{In practice, the submicron objects can move and it is important to record their subwavelength image in a non-photographic way. The most successful attempt
of such the imaging was done in \cite{NEU}, where the scanning near-field optical microscope (SNOM) whose tip moved
in a liquid medium registered the image created by a near-infrared superlens. However, in that superlens the evanescent waves did not experience amplification .
This superlens exploited the idea of the image canalization \cite{Canal}.}
In a layer of an ultimately anisotropic medium evanescent waves produced by a nearly-located object uniformly convert into propagating waves on the front interface of the layer.
On the back interface the reciprocal conversion holds and the field on the back side of the layer may reproduce the field on its front side.
It is so if the distortion of the spatial spectrum across the layer is prevented. This is possible on two conditions.
First, all spatial harmonics should propagate with nearly the same phase and group velocities. To achieve this goal, one has to engineer a proper spatial dispersion in the canalizing medium.
Second, all spatial harmonics should not experience the internal reflections at the interfaces. To achieve this goal, one has to properly select the thickness engineering the Fabry-Perot
resonance for the canalized beam.

In work \cite{NEU}, the authors managed to obtain $D\approx d=6\lambda=9.3\, \mu$m and the resolution $\delta\approx \lambda/4=390$ nm using a the layer of \emph{wire medium}.
Since in a canalizing superlens the amplification of evanescent waves is absent, it images only the objects located
in the near vicinity from the front interface, practically, at distances $\Delta=(0.05-0.1)\lambda$ \cite{Canal,Canal1,Canal11}.
Meanwhile, the total distance $D$ from the object to the image granted by canalizing superlenses can be optically large because the superlens thickness can be substantial.

Attempts to drastically increase $\Delta$ compared to $\Delta=0.1\lambda$ remained on the level of estimations -- such structures
(e.g. a superlens suggested in \cite{Min}) should have parameters which are hardly realizable in the optical frequency range.
As noticed in \cite{Canal2}, known superlenses developed for the visible and near-IR ranges could image only very closely located objects.
Thus, the canalizing superlens looks advantageous. However, the wire-medium superlens operating at the frequency $194$ THz grants the resolution
not fine enough for many applications. In order to have $\delta<100$ nm one needs to shift the operation frequency to the blue or violet bands of the visible range.
Unfortunately, in contrast to predictions of \cite{WM} a superlens presented in \cite{NEU} is not scalable to the visible range -- in this range
the wire medium is not sufficiently anisotropic \cite{WM1}.

After the publication of work \cite{STED} the stimulated emission depletion (STED) became the main topic in the literature on optical nanoimaging.
STED is a novel technique combining the dark-field microscopy and fluorescent labels attached to the object. In spite of its numerous advantages, still there are applications, especially
in the biomedicine (see e.g. in \cite{Canal2}), where the fluorescent labels in the area of the objects are prohibited. The topic of the label-free optical
nanoimaging survived in presence of STED. Therefore, works \cite{Super,Khurgin,OE,Omar} and many other similar works appeared after \cite{STED}.
Following technical solutions of a visible-range label-free superlens are currently known:
\begin{itemize}
\item
near-field superlens of silver \cite{Pen,Mel,Super};
\item
magnifying far-field superlens based on mirocavities \cite{OE,NC,Bor};
\item
canalizing (near-field) superlens based on a metal-dielectric hyperbolic metamaterial \cite{Belov};
\item
dual-metasurface near-field superlens \cite{Alitalo,OL,collective};
\item
{trinal-metasurface near-field superlens \cite{spherical}}.
\end{itemize}

The theoretical potential of the silver superlens was exhausted in work \cite{Super}, and this design solution implying two solid dielectric layers sandwiching
the perfectly flat Ag nanolayer has a very restricted area of practical applicability (namely, photographic fabrication of X-ray diffraction gratings \cite{Canal2}).
The superlens based on dielectric microcavities has also a restricted area of applications. It exploits the subwavelength size of the hot spots
corresponding to the whispering gallery resonances (WGRs) in a dielectric spherical cavity of diameter $d$ and refractive index $n$ coupled to closely located objects by near fields \cite{OE,NC}.
If the dielectric microsphere is located on a substrate covered by subwavelength scatterers and is illuminated from top, each scatterer located in the vicinity of the microsphere bottom edge
creates in it a set of hot spots corresponding to the WGRs. The number of these spots is proportional to $dn/\lambda$, their effective size is of the order
$\lambda/4n$ and their location around the sphere is determined by the location of the exciting scatterer. The techniques of \cite{OE,NC,Bor} can be considered as a development of the SNOM tip
and refers to near-field superlensing.

However, in work \cite{Hong} the far-field superlens was suggested based on the use of WGR. It was noticed that a hot spot located exactly on the opposite side of the microsphere with respect to the scatterer exciting the WGR is a source of a photonic nanojet -- a wave beam which at small distances from the sphere comprises the eavescent spatial harmonics and therefore has the subwavelength effective width. This beam propagates along the radial direction connecting the source (scatterer) the center of the sphere and the birthplace of the nanojet. Though in the far zone of the sphere (farther than the diffraction length $\Delta=d^2/4\lambda$ counted from its center) the photonic nanojet experiences the Abbe diffraction, still the subwavelength resolution in the far field is achieved. Really, if two scatterers are located on the substrate surface near the bottom edge of the sphere and the distance $\delta$ between them exceeds $\lambda/4n$ two sets of whispering gallery spots of size $\lambda/4n$ are spatially resolved inside the cavity. Therefore, two wave beams are emitted from the cavity with the angle $\phi=(\lambda/4n)/(d/2)$ between their directions. At the distance equal to the diffraction length their centers are distanced by $d/8n$. For high-order WGRs $d>8\lambda n$. Since at the distance from the cavity equal to the diffraction length the effective width of the photonic nanojet is nearly equal $\lambda/2$, the spatial separation of two nanojets exceeds their effective width at least twice. So, two point-wise scatterers separated by a subwavelength gap, create in their far zone two spatially resolved wave beams which magnify the image of this dimer of scatterers. The magnification is sufficient to record the subwavelength image by conventional optical detectors. e.g. by a CCD camera. The lateral resolution $\lambda/4n=\lambda/6$ granted by a glass sphere fabricated with a nanometer precision was experimentally verified in \cite{Hong}. The aforementioned physics of the imaging was understood in work \cite{Yang}. Also, in this work the lateral spatial resolution was further enhanced replacing the glass by a medium with higher refractive index.

However, in this and further works exploiting the effect the authors have not explained why the nanojet generated by a scatterer is emitted from the opposite extremity of the cavity.
To our opinion, the physics of this imaging is the same as the physics of directional radiation from a notched optical cavity at frequencies of a high-order WGR modes. This radiation was revealed in work \cite{Boriskina}. The subwavelength notch, similarly to a closely located scatterer, cancels the uncertainty in the location of the WGR hot spots. Simultaneously the near-field coupling opens a channel for leaking the WGR eigenmode from the cavity. If an eigenmode of a cavity in absence of losses is ideally decoupled with propagating waves of the ambient, the quasi-mode of a defective cavity is linked to them at two points -- that of the notch and that located on the opposite side of the cavity. This way, the notch becomes a source of a directional wave beam which is generated on the opposite side of the cavity \cite{Boriskina}. Notice, that before this phenomenon turned out to be useful for nanoimaging, it served for creation of high-directionality microlasers (see e.g. the seminal paper \cite{WGRlasers}). Finally, it worth noticing that the phenomenon of the nanojet generation in the perturbed WGR cavity grants not only lateral but also 3D resolution. Two scatterers located on the same radial line of a microsphere and separated by a subwavelength gap, evidently send their photonic nanojets in the same direction. However, the phases of these wave beams  are different. Therefore, the birthplaces of two beams can be resolved involving the so-called Mirau interferometry. This interferometric spatial resolution is even finer than the lateral one \cite{Kassamakov1}.

These spherical superlenses may form an array and can be moved on the surface so that to gradually image the whole area, as it was suggested in \cite{Kassamakov1}. However, they do not completely resolve the issue of far-field label-free subwavelength imaging. Really, the microspheres create the image in the reflected light and are located directly on the substrate. Thus, they may mechanically interact with the subwavelength objects and disturb them. In this paper, we aim a remote subwavelength image -- that created by the transmitted field.

A canalizing hyperbolic metamaterial of alternating silver and dielectric nanolayers suggested in \cite{Belov} seemed in 2006 to be the most promising solution for a superlens operating with  transmitted waves. However, since that time comprehensive studies have shown that the impact of the plasmonic dissipation in Ag restricts the allowed total thickness of such the superlens \cite{Canal11}. In \cite{Rafal} the maximal allowed thickness of the superlens $d=D-2\Delta$ corresponding to the
spatial resolution $\delta\approx \lambda/10$ was predicted as $d\approx 0.9\lambda$. In \cite{Khurgin} this calculation was refined and the maximal number $M=20$ of stacked silver-dielectric bilayers offering this resolution was obtained. The restriction $M\le 20$ reduces the allowed $d$ to $400-500$ nm. If we target the resolution $\delta\approx \lambda/5$ the allowed thickness $d$ increases twice and does not exceed $1\, \mu$m.

However, there is another, more severe, limitation of the allowed thickness $d$ for a hyperbolic-medium superlens. This is the roughness of nanolayers
which accumulates along with their growing number $M$ \cite{ALD}. In practice, to obtain $M=20$ with the needed precision is very difficult. Therefore, the research groups capable to reproduce the results of \cite{ALD} have concentrated on a tapered version of this superlens that looked to be more practical. This seemingly more practical hyperbolic-medium superlens
is a magnifying superlens, usually called hyperlens \cite{FF1,FF11}. The image magnification presumably allows to keep the subwavelength details of the object in the far-field image.
Both cylindrical (granting the 1D magnification) or spherical (2D magnification) hyperlenses were developed (see e.g. in \cite{FF2,FF3}).
However, the cylindrical hyperlens offers the subwavelength resolution along one line only. As to the spherical hyperlens, grating the 2D subwavelength resolution 
the practical limit is $M=9$. It grants a very modest resolution $\delta\approx \lambda/3$ and only a twofold magnification.
Moreover, a spherical hyperlens with nine metal-dielectric bilayers is a submicron device whose object area diameter is smaller than 200 nm \cite{FF4}.
The hyperlens turns only a slight (though very expensive) improvement of the well-known dielectric SNOM tip.

As to dual-metasurface optical superlenses, they were developed insufficiently. Their advantages were not clarified in \cite{Alitalo,OL}. A trinal-metasurface superlens presented in \cite{spherical} is an extent of the dual-metasurface superlens which grants a higher intensity in the image plane (for the same resolution). However, this improvement implies a more expensive fabrication. It is not surprising, that these techniques have been nevr mentioned in popular overviews of subwavelength imaging (see e.g. in \cite{Canal2} and \cite{SLO}).
{In this paper, we present a} {theoretical and numerical} {study of a dual-metasurface optical superlens based on both novel semi-analytical model and full-wave simulations.
We discuss the shortage of the precedent works and show that} {the new model} {allowed us to obtain much better results.
Finally, in Appendix we describe a novel application of a dual-metasurface superlens actual for biological nanoimaging and suggest an affordable fabrication technique based on the self-assembly.}

\section{Simplistic model of a dual-metasurface superlens}

\begin{figure*}[t]
 \centering
 \includegraphics[width=0.8\textwidth]{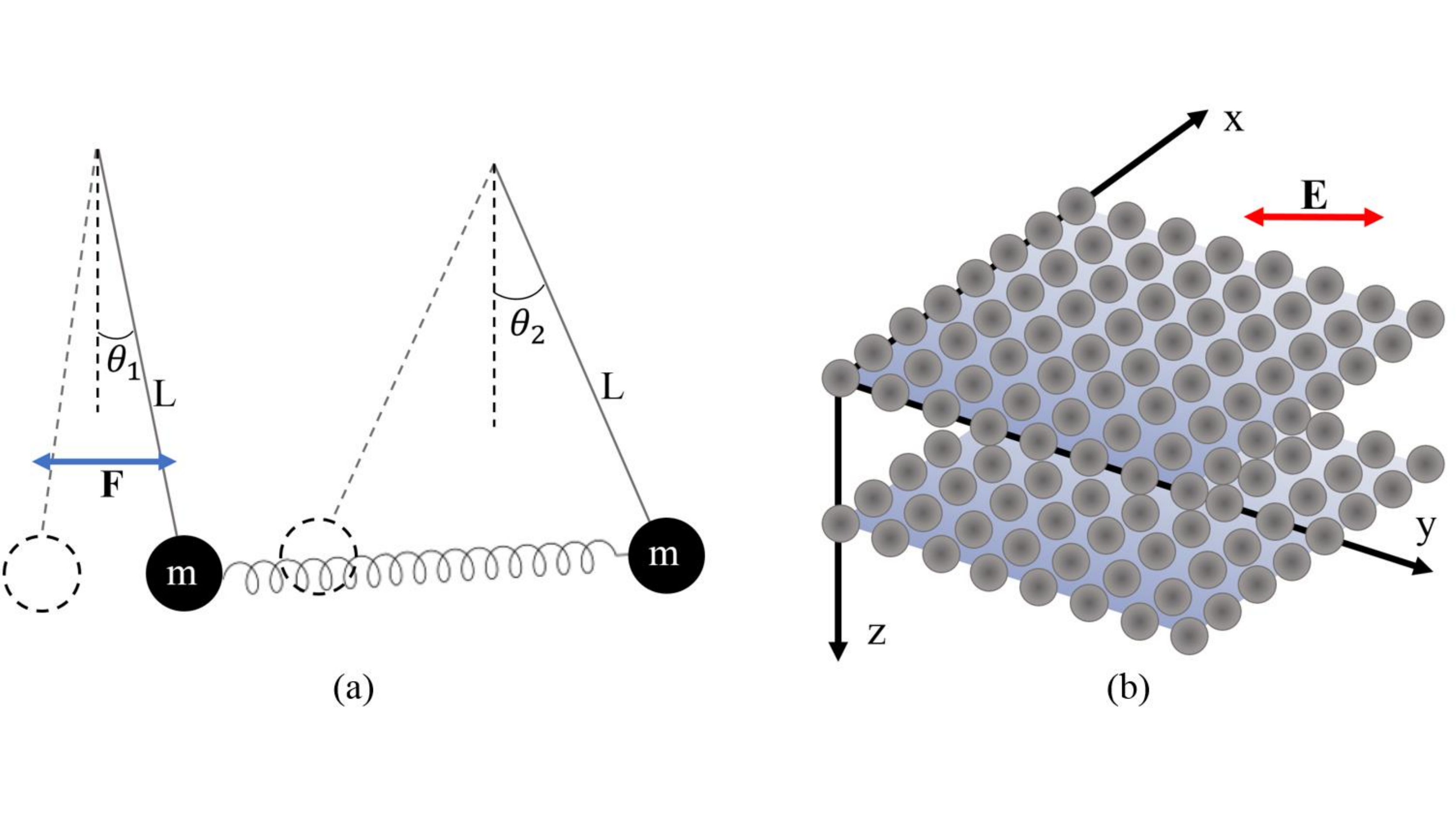}
 \caption{Conceptual illustration of the underlying physics: (a) -- a dimer of two elastically coupled pendulums excited by an oscillating external force $\bf F$;
 (b) -- a dimer of two plasmonic metasurfaces excited by a hot spot of light with electric field $\bf E$.}
 \label{fig1}
\end{figure*}

In the course of classical mechanics (see e.g. in \cite{CM}) the following effect is well known. In a system of two identical oscillators
such as two elastically coupled pendulums shown in Fig.~\ref{fig1}(a), the resonance of a single pendulum symmetrically hybridizes. However, if
we excite one pendulum by an external force $\-F$ oscillating with the resonance frequency  $\o_0$ of the single pendulum, the amplitude of the
oscillations in the back pendulum (in Fig.~\ref{fig1}(a) it is denoted $\theta_2$) will be much higher than that of the front one (in Fig.~\ref{fig1}(a) -- $\theta_1$).
In fact, $\theta_2/\theta_1$  -- amplification of the oscillation across the dimer -- is maximal at $\o_0$ and holds for whatever frequency
located in between the normal modes $\o_1$ and $\o_2$. At both these frequencies $\theta_2=\theta_1$ and the dimer oscillates as a solid pendulum. At frequencies $\o>\o_1$ and
$\o<\o_2$ the front pendulum oscillates stronger ($\theta_2<\theta_1$).

In work \cite{Mas} it was assumed that a TM-polarized surface wave (polariton) in a resonant metasurface is analogous to a lumped Lorenztian oscillator such as a pendulum of Fig.~\ref{fig1}(a).
If so, a dual-metasurface structure should be analogous to two coupled pendulums because the near-field coupling can be
inductive for properly chosen gap $h$ between them, and the inductive coupling of two circuits is effectively elastic.
This brilliant guess has been validated by a microwave experiment where an array of wire meanders supported a polariton which hybridized for a dimer of two parallel arrays.
The electric field intensity was measured in both front and rear array excited by a point source oscillating at the frequency of the individual resonance
$\o_0$. For a properly chosen distance $h$ between the arrays the intensity in the rear array turned out to be much higher than that in the front one.
The growth of the evanescent waves across the gap $d$ between two metasurfaces was also measured and, since these theoretical expectations were confirmed,
the subwavelength imaging in the dual-metasurface structure was predicted \cite{Mas}. This prediction was experimentally confirmed
in work \cite{Mar} where a more elaborated microwave constituent (a split-ring resonator instead of a meander) was used for in the metasurfaces.

Implementation of this idea in the visible range was suggested in work \cite{Sim1}, where the dispersion of a plasmon polariton in a chain of Ag nanospheres
was theoretically studied. It was shown that the regime where this polariton is qualitatively similar to an eigenmode of a lumped oscillator (all wave numbers
correspond to a very narrow band of frequencies) is feasible. Since the polariton wave package propagating along the chain mimics
the pendulum, two parallel chains should grant a 1D subwavelength imaging similar to that obtained in \cite{Mar}. Further, it was shown that the similar dispersion
can be engineered in a planar (square) array of the same nanospheres. In work \cite{Alitalo} we theoretically obtained the subwavelength resolution of two point dipoles distanced by $\delta=0.34\lambda$, whereas the distance between the object and image planes was $D=h+2\Delta=0.55\lambda$ (nearly 240 nm).
We concluded that the dual metasurface depicted in  Fig.~\ref{fig1}(b) can be really considered as a distributed analogue of a classical dimer of identical oscillators with elastic coupling.

\begin{figure*}[t]
 \centering
 \includegraphics[width=0.8\textwidth]{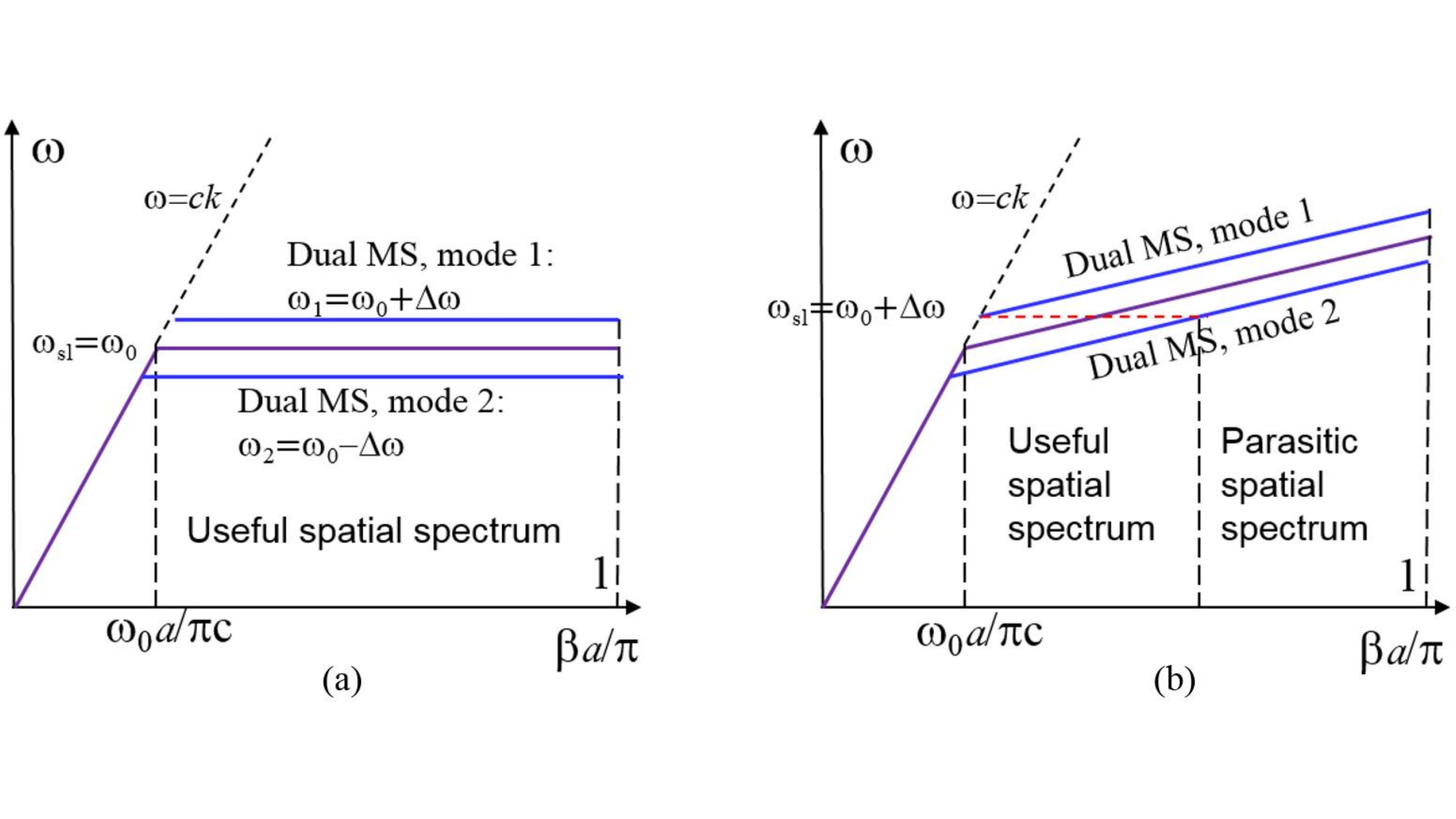}
 \caption{(a) Ideal (target) dispersion of a single metasurface and that of a dual-metasurface structure.
 (b) Dispersion of a realistic metasurface with the forward wave and that of a realistic dual-metasurface structure.
 Bands of spatial frequencies corresponding to both useful and parasitic contributions into the image are shown for the optimal
 operation frequency $\o_{sl}$ in both cases (a) and (b).}
 \label{fig2}
\end{figure*}

In works \cite{Syd,Syd1}, the role of the dispersion of a single metasurface was explained in details. Though these theoretical and experimental studies were done
for an array of split-ring resonators, some statements of these works are applicable to whatever dual metasurface.

The dispersion of an idealized dual-metasurface superlens in shown in Fig.~\ref{fig2}(a). The range of spatial frequencies $\beta a/\pi<\o_0a/\pi c $
($a$ is the internal period of the metasurface, $\beta$ is the wave number) corresponds to propagating waves -- light line $\omega=ck$ in the plot.
The range $\o_0a/\pi c<\beta a/\pi< 1$ limited by the bound of the Brillouin zone of the metasurface is that of the guided wave -- polariton.
The low-frequency polariton mode of the single metasurface in the ideal case has the group velocity that equals identically zero for all $\beta>\o_0/c$. In the dual metasurface
this mode splits onto two normal modes 1 and 2. If the coupling is effectively elastic, the dispersion of this hybridized polariton keeps basically the same as that of the generic one, and modes 1 and 2 keep the zero group velocity. Therefore, for any $\beta$ in the polariton range the frequency $\o_0$ is in the middle of the eigenfrequencies $\o_{1,2}$ of the normal modes. Such the idealized dual-metasurface superlens ideally mimics the dimer of pendulums. At frequency $\o_0$ the whole band of the guided spatial frequencies is a useful spatial spectrum contributing into the near-field image. In other words, all evanescent waves oscillating at $\o_0$ and impinging the front metasurface will experience the same maximal amplification across the gap $h$. This ideal superlens should operate at frequency $\o_{sl}=\o_0$, that grants the maximal ratio $D/\delta$, and minimal $\delta\ll \lambda$.

Of course, in reality, the group velocity of the polariton cannot be identically equal zero. A realistic dispersion may correspond either to a forward wave, or to a backward
wave, or to a wave which is backward within one band of $\beta$ and forward within another band and pass through the point at which $\partial \o/\partial \beta=0$ (see more details in \cite{Sim1}). In Fig.~\ref{fig2}(b) we schematically show the case of the forward eigenwave.
Here, it is reasonable to choose the superlens operation frequency $\o_{sl}=\o_1$, where $\o_1$ is the eigenfrequency of mode 1 corresponding to the point
where mode 1 diverts from the light line (notice, that in work \cite{Syd1} the eigenwaves were backward, and the best image was obtained choosing $\o_{sl}=\o_2$.).
With this choice of the operation frequency, the useful spatial spectrum is maximally broad.
However, the amplification of evanescent waves across the superlens is non-uniform -- maximal only in the middle between modes 1 and 2 (where the red dashed line crosses the solid violet one).
For smaller and larger $\beta$ in between the modes 1 and 2 the amplification of evanescent waves is lower. This non-uniformity results in a distortion of the image. Moreover, mode 2 is excited at $\o_{sl}$ that means the resonant enhancement of the near field in the whole structure i.e. implies a parasitic image of the superlens itself. Spatial frequencies which are not amplified across the superlens also contribute into this parasitic image because the corresponding evanescent waves experience the reflections at the constitutive metasurfaces of the superlens.
These implications of the non-zero group velocity of the polariton were understood by the authors of \cite{Syd1}.

{Since the birefrequence $\Delta\o$ is inverse proportional to the gap $h$ \cite{Syd1}, when $h>\lambda/4$, $\Delta\o$ is small and} the useful range of spatial frequency is very narrow and the image of a strongly subwavelength source (a circular loop of the diameter $\lambda/10$) was not detectable on the background of the parasitic image. The situation
similar to that schematically depicted in Fig.~\ref{fig2}(b) was observed in \cite{Syd1} when $h=\lambda/5$. Then $\Delta\o$ was large enough,
and the band of useful $\beta$ was broad enough. Therefore, in the image plane distanced by $D=0.4\lambda$ from the source plane the image of the loop had the size $0.3\lambda$
and could be detected. The best imaging operation corresponded to $h\le \lambda/10$, when the gap $2\Delta\o$ between modes 1 and 2 was so wide that the
operation frequency did not intersect with the dispersion curves of the eigenmodes 1 and 2. In other words, in this case the useful range of spatial frequencies
occupied the whole Brillouin zone like it holds in Fig.~\ref{fig2}(a). Though the image of the radiating loop was distorted (due to the non-uniform amplification of the evanescent waves)
its size was equal to the size of the object -- $\lambda/10$ that left no doubts in the possibility of deeply subwavelength resolution.
The source was located at the distance $\Delta=\lambda/20$ from the front metasurface, and the distance form the object to the image
was $D=h+2\Delta=\lambda/5$. A so small distance was the cost of the poor dispersion of the constitutive metasurface in the range of surface waves.
In order to increase $h$ and $D$, one has to engineer a polariton with a much smaller group velocity.

This insight gave a pulse for the improvement of the plasmonic superlens. In a planar array of silver nanospheres, the rather flat dispersion
of the surface plasmon-polariton  (SPP) corresponds to the normal (vertical) polarization of the nanospheres, whereas the horizontal (tangential) polarization is a forward wave with
substantial group velocities in the guided wave region. In \cite{Alitalo} we considered the dual metasurface depicted in Fig.~\ref{fig1}(b) from
Ag or Au nanospheres in PMMA ($\va=2.3$). In our calculations, the nanospheres were modeled by the point dipoles and the model of calculations was semi-analytical since the problem formulation allowed us to use the uniform-space Green function for calculating the dipole field. In the structure of Ag spheres, the frequency $\o_{pr}$ -- plasmon resonance of a single nanoparticle -- referred to the blue range, and $\o_{sl}$ was in the green band. For both - silver and golden -- superlenses the optimal parameters were found from the analysis of the dispersion of a single metasurface. For that of Ag spheres the dispersion curve of the vertical polarization in the region $\beta>\o/c$ was maximally flat when the particle diameter $d$ was equal $56$ nm and the grid period $a=65$ nm. In that work only the dispersion of the single metasurface was calculated, the question on the birefringence was not answered. It was assumed that the dispersion curves of the vertical polarization plotted for a single silver and golden metasurfaces in Figs. 2 and 3 of this work split onto similar curves for a dual metasurface. The optimal distance $h=130$ nm between two arrays was specified in a set of semi-analytical simulations. The superlens frequency $\o_{sl}$ was slightly higher than $\o_0$ and this difference was treated as an impact of losses since the result was robust with respect to the number $N$ of unit cells along both $x$ and $y$ axes). Really, calculations of the dispersion diagram were done for $N\rightarrow \infty$ and
optical losses in Ag and Au were neglected in order to make the problem periodical along $x$ and $y$. Meanwhile, these losses were taken into account in the simulations of the electric field
intensity for the superlens with finite $N$ excited by the sources to be imaged.
{The role of losses is destructive for subwavelength imaging due to a significant restriction of the maximal value of $\beta$ corresponding to the surface plasmons compared to $\pi / a$. The dispersion of a plasmonic metasurface (single and dual) practically does not suffer of losses in the  region $\beta a <\pi/2$, however
in the region $\beta a \rightarrow  \pi$ becomes weird and the group velocity loses the physical meaning there. The same effect holds for whatever SPP, e.g. in solid silver layers \cite{nanoplasmonics}.}

However, the difference $\o_{sl}-\o_{0}$ had a different reason, the same as that of a modest result obtained for the best (Ag) optimized superlens: $\delta=0.34\lambda$ for $D=0.51\lambda$.
A so modest spatial resolution at a half-wave distance was caused by the dispersion of the dual-metasurface drastically different from that of the single one.
Only in a single metasurface the eigenmodes with the vertical and horizontal polarizations may be excited separately.
For a dual-metasurface these polarizations hybridize and the dispersion diagram is not similar to those of Figs.~\ref{fig2}(a) and (b).
The use of only vertically polarized sources in \cite{Alitalo} does not allow to reduce the impact of the horizontal polarization since the dispersion of the dual metasurface is its intrinsic characteristics and does not depend on the source.

In \cite{OL} we suggested to use a metasurface of oblate Ag spheroids. In this case the hybridization of the vertical and horizontal polarizations can be suppressed.
In an array of oblate spheroids (nanotablets) the resonance frequency $\o_0$ turns different for the horizontal and vertical polarizations. The vertical dipoles in a single metasurface resonate at a lower frequency and the horizontal dipoles at this frequency are negligibly small. Then the dispersion curve of the single metasurface really splits onto tow parallel curves for
the dual metasurface like those depicted in Figs.~\ref{fig2}(a) and (b). However, the flatness of the dispersion curve
is worse for a single planar array of oblate spheroids compared to that of spheres. Therefore, in this way we achieved only a slight increase of $D$ from $D=0.55\lambda$ to $D=0.63\lambda$ and a slight improvement of the resolution from $\delta=0.34\lambda$ to $\delta=0.31\lambda$. Notice, that these calculations, as well as calculations of \cite{Alitalo}, were approximate. Silver nanotablets were modelled by point dipoles located in a uniform host and objects to be resolved were not scatterers but vertically polarized sources.

{In the present paper, we use a novel approach to a dual-metasurface superlens rejecting both simplistic model of \cite{Alitalo}
and suppression of the transverse polarization suggested in \cite{OL}. This new approach results in a qualitative improvement of our superlens.}

\section{Preliminary simulations of the single and dual metasurface}

\begin{figure*}[t]
 \centering
 \includegraphics[width=0.9\textwidth]{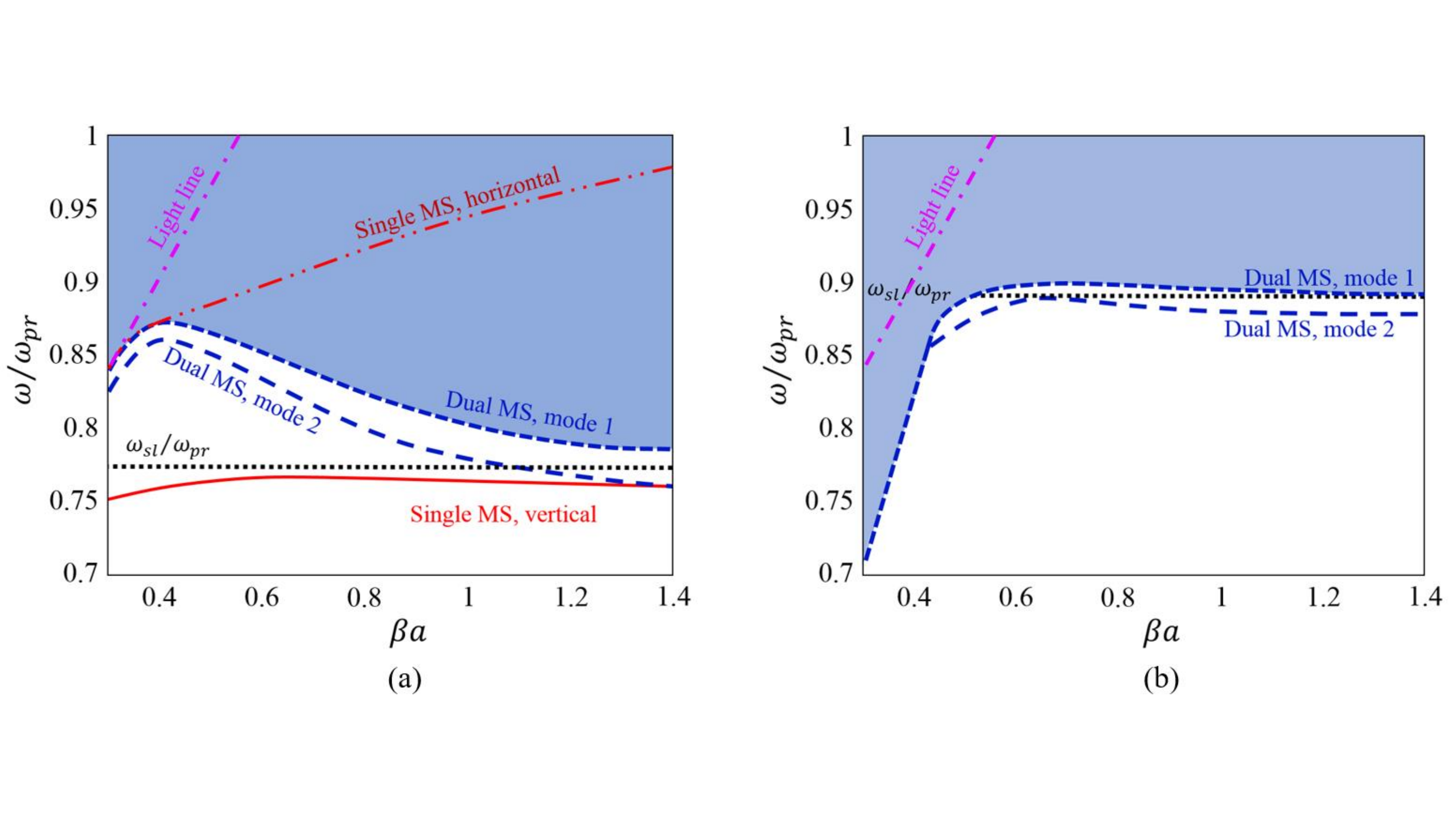}
 \caption{Dispersion of the dual-metasurface superlens in a homogeneous host medium (point-dipole model of nanospheres {while $\o_{sl}$ is superlens operation frequency and $\o_{pr}$ is plasmon resonance of a single nanoparticle}).
 (a) High-frequency passband with its $\beta$-dependent lower bound and the low-frequency dispersion curve for a dual metasurface with parameters suggested in \cite{Alitalo}.
 The dispersion curves calculated in this work for a single metasurface (vertical and  horizontal polarizations) are not relevant for the superlens. The spectrum of evanescent waves
 experiencing the amplification at the superlens frequency ($\beta a=1.1\dots \pi$) is significantly narrower than that predicted in \cite{Alitalo}.
 (b) The similar dispersion diagram for improved design parameters compared to \cite{Alitalo}. In this case, frequency $\o_{sl}$ does not cross the dispersion curves
 of the superlens.}
 \label{fig5}
\end{figure*}

Preparing the paper \cite{OL} we developed a Matlab code that calculates the dispersion diagram of eigenwaves in a periodic ($N\rightarrow \infty$)
dual-metasurface structure. This code resolves the general dispersion equation of a periodic array which follows from the tensor equation 
${\=\alpha(\o)}^{-1}\cdot \-p=\=\sigma(\o,\-\beta)\cdot \-p$ describing any periodic system of dipole particles (see e.g. in \cite{Shore}). Here $\=\alpha$ is the particle polarizability (in the case of spheroidal particles it is a uniaxial tensor) and $\=\sigma$ is the interaction factor of the periodic array. A square array (monolayer or bilayer) is nearly isotropic in plane $(x-y)$, and it is enough to consider the eigenwaves propagating along one of the array axes, e.g. $\beta=\beta_x,\, \beta_y=0$. The dipole moment $\-p$ of the reference particle of an array is polarized in the $(z-x)$ plane that corresponds to TM-polarized eigenmode. In this case, the local-field equation is a system of two scalar equations for $p_x$ and $p_z$ (reference dipoles $\-p$ in both planar arrays of the dual metasurface are identical). The interaction factor a dual square array of electric dipoles is a dyad with three nonzero components $\sigma_{zz},\sigma_{xx}$ and $\sigma_{zx}=\sigma_{xz}$. The expressions for these components in a form of converging series correspond to the special case $N_a=2$ of the dynamic dipole sums obtained in work \cite{Shore}. Formulas (120-124) of this work describe the components of the interaction dyad $\=\sigma$ for an equidistant set of $N_a$ planar arrays of identical dipole particles.

In the present case (for the spheres), $\=\alpha$ degenerates to a scalar $\alpha$. It is not a trivial static polarizability of the sphere -- the inverse polarizability $\alpha^{-1}$ entering the local-field equations comprises the scattering losses of an individual particle \cite{Alitalo}. Eigenwaves correspond to the vanishing determinant of the local field equations. Thus, we come to the dispersion equation of a dual metasurface:
$$
\left(\alpha^{-1}-\sigma_{xx}\right)\left(\alpha^{-1}-\sigma_{zz}\right)=\sigma_{xz}^2.
$$

Numerical solution of this equation in Matlab has shown that the dispersion of the dual metasurface of nanospheres is drastically different from two
curves depicted in Fig.~\ref{fig2} even if the dispersion curve of the single one is almost ideally flat. In other words, for nanospheres the simplistic model is not applicable. 
Our Matlab code calculates in the coordinates $(\beta-\o)$ an area free of eigenfrequencies. We call it the bandgap between modes 1 and 2. Here mode 2 is described by a single dispersion curve because it is the wave with dominating vertical polarization. As to mode 1, it is in fact, a set of intersecting dispersion curves that occupy a certain area in our dispersion diagram. For our purposes only the lower bound of this area is important because it determines the bandgap. Notice, that on this dispersion diagram we do not show the range $\beta a=1.4\dots \pi$ because for high spatial frequencies a small error in the interaction factor of the superlens causes a huge error for the eigenfrequency. Our semi-analytical model practically does not work for wave numbers $\beta$ approaching to the bound of the Brilouin zone. However, for our purposes the area of the dispersion diagram shown in Fig.~\ref{fig5} is sufficient.

In Fig.~\ref{fig5}(a) we depict the dispersion of the single metasurface calculated in work \cite{Alitalo} and can see that both these curves -- that corresponding to the vertical 
polarization and that corresponding to the horizontal one -- are not very irrelevant for the superlens operation. This means that the optimization performed in \cite{Alitalo} was 
not fully correct. On one hand, the dispersion curve of a single metasurface for the vertical polarization passes very closely to the optimal operation frequency $\o_{sl}$
shown in Fig.~\ref{fig5}(a). This frequency was estimated in \cite{Alitalo} correctly. On another hand, the useful range of spatial frequencies, as it is seen 
on this diagram, starts at $\beta a=1.18$. This interval is, therefore, rather narrow compared to the whole band of guided waves $\beta a=0.3\dots \pi$. Since only a rather narrow spatial spectrum of evanescent waves is amplified across such the superlens, it is not surprising that the results of \cite{Alitalo} were modest. 

\begin{figure*}[t]
 \centering
 \includegraphics[width=0.9\textwidth]{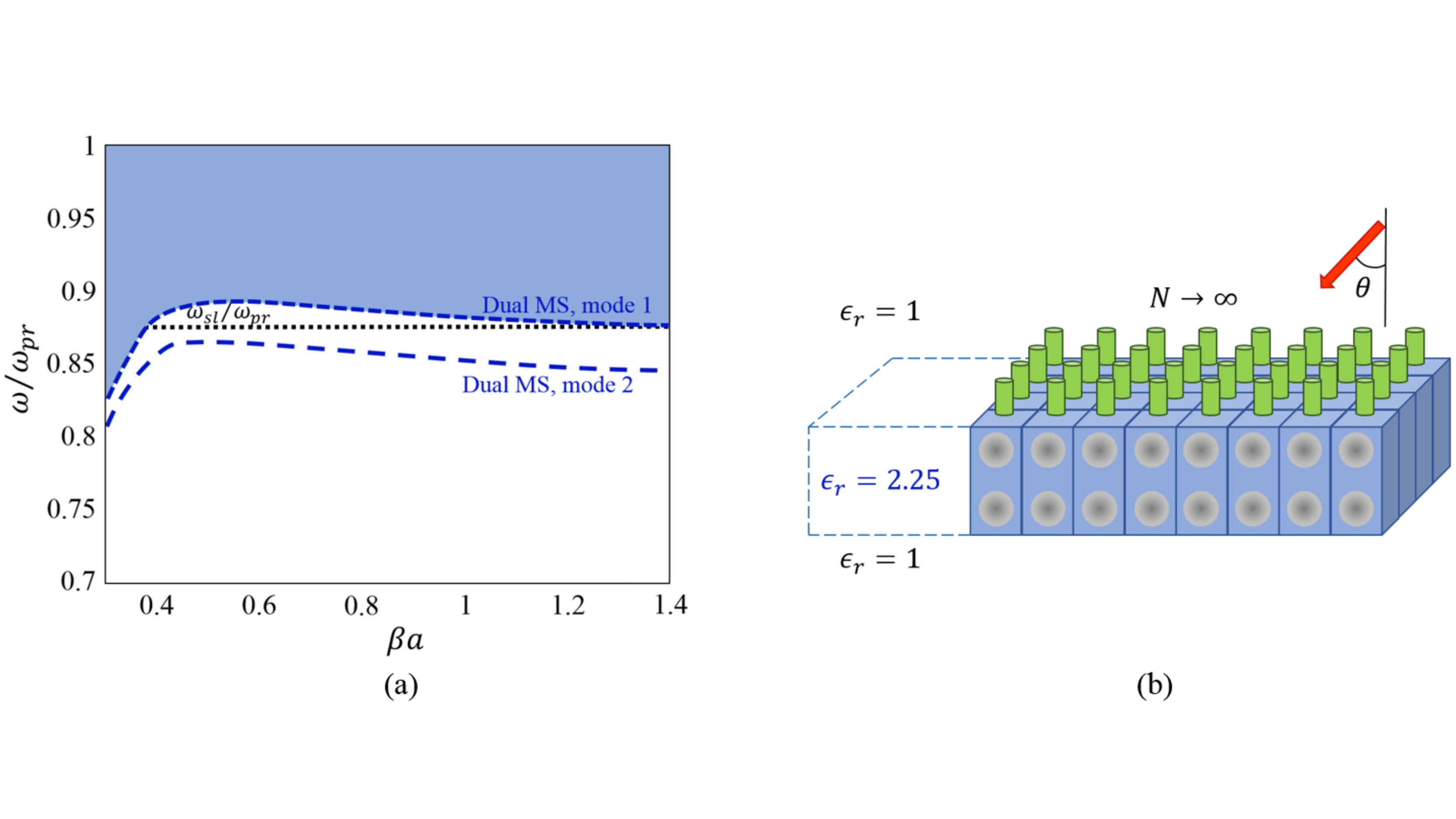}
 \caption{(a) Dispersion diagram of the dual-metasurface superlens in a homogeneous host medium $\va=1.625$. Point-dipole model of nanospheres is used.
 (b) Sketch of the structure for which the full-wave numerical simulations of the plane-wave incidence were done.}
 \label{fig6}
\end{figure*}

Only a slight improvement was obtained in \cite{OL}, where the horizontal polarization was suppressed, because the nanotablets replacing the 
spheres do not grant the flatness of the dispersion curves achievable for an array of nanospheres. Therefore, in the present paper we neither rely on the dispersion of a single metasurface nor 
try to damp the horizontal polarization. We simply perform a set of Matlab calculations of the dual-metasurface dispersion aiming to engineer a bandgap
between the modes 1 and 2 in the maximally broad range of guided spatial frequencies $\beta$. Maximal interval $\beta a=0.5\dots \pi$ was obtained in 
the case $\va=2.3$ (as in \cite{Alitalo}) choosing the diameter of nanospheres $d=60$ nm, the grid period $a=80$ nm and the distance between the centers of two planar grids $h=130$ nm. 
The dispersion diagram for this case is shown in Fig.~\ref{fig5}(b). The superlens operation frequency $\o_{sl}$ centers this bandgap. Comparing Figs.~\ref{fig5}(a)and (b)
we see that the geometric parameters of the superlens suggested in \cite{Alitalo} ($d=56$ nm and $a=65$ nm for $h=130$ nm) were not very close to optimal ones.
Parameters of silver were taken from \cite{Ag} as well as in \cite{Alitalo,OL}.

Next, let us repeat this optimization for a more realistic permittivity of the ambient. Since our semi-analytical model
does not allow us to take into account the presence of the interfaces, we replace the layered ambient by a uniform ambient
whose effective permittivity is equal $\va=(\va_{\rm glass}+1)/2=1.625$.
Keeping the same geometric parameters, as we have engineered for $\va=2.3$ we obtain a broader bandgap between the modes 1 and 2 -- we can see it in Fig.~\ref{fig6}(a).
The choice of the superlens frequency shown in this plot as a black dashed line gives a hope to the higher amplification of evanescent waves and lower distortion of the image
compared to the case when $\va=2.3$. This expectation will be confirmed below by full-wave simulations.

In order to estimate the impact of the fabrication tolerances, we calculated a set of dispersion diagrams varying both $d$ and $a$.
For the case $\va=1.625$ the allowed deviations are reasonably large. If $a=80$ nm the complete bandgap for surface waves corresponds to $d=57-70$ nm,
if $a=70$ nm the same refers to $d=55-64$ nm. A pessimistic estimate for the tolerances is $a=75\pm 5$ nm, $d=60\pm 4$ nm.
These calculations were done within an approximate model where the dispersion of an infinite ($N\rightarrow \infty$) superlens
was calculated for a bilayer array of point dipoles in a uniform host medium. However, full-wave simulations using the commercial software CST Studio have confirmed the conclusions referred to the tolerances.

A set of full-wave simulations of the infinite superlens refers to preliminary calculations since
on this stage we did not aim the imaging. The goal was to confirm the predictions of the semi-analytical model
and the bandgap between the SPP regions in the optimized infinitely extent $N|rightarrow\infty$ superlens. The simulated structure is illustrated by Fig.~\ref{fig6}(b).
On top of the superlens a periodic array of scattering objects was located. Since the SPP is TM-polarized, the incident plane wave was TM-polarized, as well.
Due to the translation invariance, the problem depicted in Fig.~\ref{fig6}(b) reduces to the cell problem with the $a\times a$ unit cell (in the horizontal plane,
in the vertical direction the cell is restricted by the radiating surfaces). This problem formulation drastically reduces the computation time and resources.
We studied this infinite superlens with and without the scattering objects. Without the objects, the eigenmodes of the superlens if excited
cannot be easily recognized in the field maps calculated by the solver. However, SPPs should be excited by a plane wave within the consolidated plasmon resonance band of the dual metasurface
due to the effect of Wood's anomalies \cite{Wood}. Really, inspecting the field intensity maps we see the sets of hot spots varying with
the incidence angle and observe the angle-dependent local maxima of absorption, corresponding to the appearance of the hot spots.
This study allowed us to specify the bands of the high-frequency SPPs (mode 1) and low-frequency SPPs (mode 2) and the bandgap between them
at 646-649 THz where the absorption has a local minimum.

Notice, that our analytical point-dipole model for a uniform host medium with $\va=1.625$ predicted the SPP for the single metasurface at 655-660 THz
and the bandgap at nearly the same frequencies for the dual metasurface. In another word, the full-wave simulations fit our analytical model
with amazingly small discrepancies 1-2\%. However, even in the frequency range of the superlens in absence of the objects still we observe some (not very high) local field enhancement near the constitutive nanospheres. In our point-dipole model, local field enhancement within the bandgap was absent. {Also, full-wave simulations pointed out another optimal distance ($h=120$ nm instead of $130$ nm) between the centers of the top and bottom arrays.

The most important result of these simulations was the evidence of the evanescent waves amplification between two metasurfaces within the SPP bandgap. This amplification is absent beyond this range. To earn the evanescent waves, in our simulations, we used 1) dielectric objects giving the low contrast in the intensity plots, 2) the silver objects granting the highest contrast but also brightest parasitic hot spots, and 3) perfectly electrically conducting (PEC) objects. The PEC scatterers produced a sufficient contrast and modest parasitic hot spots and were the most convenient sources for observing the amplification of the evanescent waves across the superlens. However, simulations of the periodic structure depicted in Fig.~\ref{fig6}(b) were only a preliminary step, which proved the predictive power of our semi-analytical model. Below we report the full-wave simulations of more practical structures, whose design parameters were found in our semi-analytical calculations.

\section{Full-wave simulations of a superlens}

In Fig.~\ref{fig7}(a) the sketch of a finite-size superlens illuminated by a TM-polarized plane wave is presented. In this example, it comprises $N_x\times N_y$ unit cells, where
$N_x=7$ and $N_y=5$. We increased $N_x$ and $N_y$ from 5 to 28, and clearly saw that for $N_{x,y}>7$ the imaging of the objects {(cylinders and spheres made of PEC, silver or dielectric with relative permittivity $\va=3$)} located in the central area of the superlens keeps independent on the superlens size. For certainty, below we depict the results of the simulations for $N_x=N_y=N=9$. Notice, that the bi-sectorial incidence plane is favorable for seeing the non-distorted images. At the edges of the superlens intersecting the incidence plane there are parasitic hot spots. If the incidence plane is bi-sectorial these hot spots are located at the corners i.e. are maximally distant from the image area.

\begin{figure*}[t]
 \centering
 \includegraphics[width=0.9\textwidth]{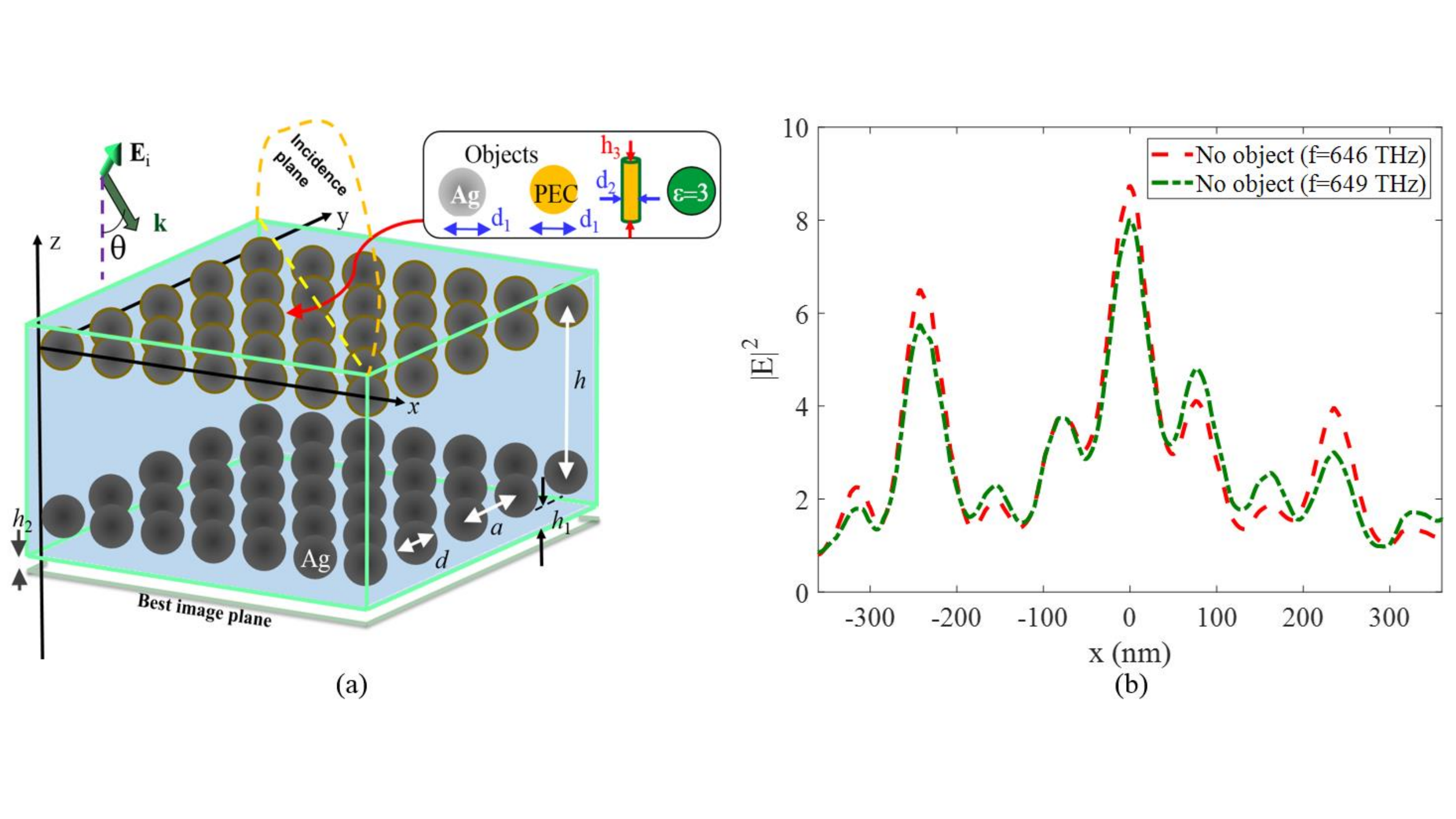}
 \caption{(a) A finite-size superlens with $N_x\times N_y=35$ unit cells illuminated by a plane wave in bi-sectorial incidence plane under incidence angle $\theta = \phi = \pi/4$
 in presence of one or more objects.  (b) The normalized intensity profile calculated on the rear surface of the superlens
 $z=-h-d/2-h_1$ along the line $y=N_ya/2$ for $N_x=N_y=9$ at two operation frequencies (646 and 649 THz).}
 \label{fig7}
\end{figure*}

Though we eliminate these brightest parasitic hot spots from the image area, still a parasitic intensity enhancement in absence of the objects holds.
In Fig.~\ref{fig7}(b) we depict the local intensity distribution along $x$ for $y=Na/2$ at the back side of the superlens. This back side is the plane $z=-h-d/2-h_1$, where
$h_1=(a-d)/2$. The best image plane is shifted with respect to the superlens back side by a value $h_2$ that depends on the size of the object.
In our simulations, we locate one or two scattering nanoobjects on the front side of the superlens, as it is shown in Fig.~\ref{fig7}(a), and
$h_2$ is nearly equal to one half of the vertical size of the object. However, the variation of the local intensity profile along the axis $z$
is not very sharp in the rear area and for small objects ($d_1\le 60$ nm and $h_1\le 60$ nm) one may approximately adopt that the best image plane
is the back interface of the superlens.

In Fig.~\ref{fig7}(b) the intensity is normalized to that of the incident wave and presented for two frequencies 646 and 649 THz -- bounds of the presumed superlens operation band.
In accordance to the Airy formula for the infinite glass layer of the same thickness $h+d+2h_1=h+a=200$ nm at the back side of the plate illuminated by the plane wave with the unit electric amplitude $|E|^2\approx 0.8$. We observe a significant enhancement of the intensity compared to this value in the area of the plate ($|x|<360$ nm) that is not the effect of the finite size
of the superlens. The same enhancement was obtained in the simulations of the periodic structure and reported above.
This enhancement is much higher at the frequencies of SPPs, but even in the bandgap it is not a negligible effect, and
the image cannot be observed without this parasitic background.

\begin{figure*}[t]
 \centering
 \includegraphics[width=0.9\textwidth]{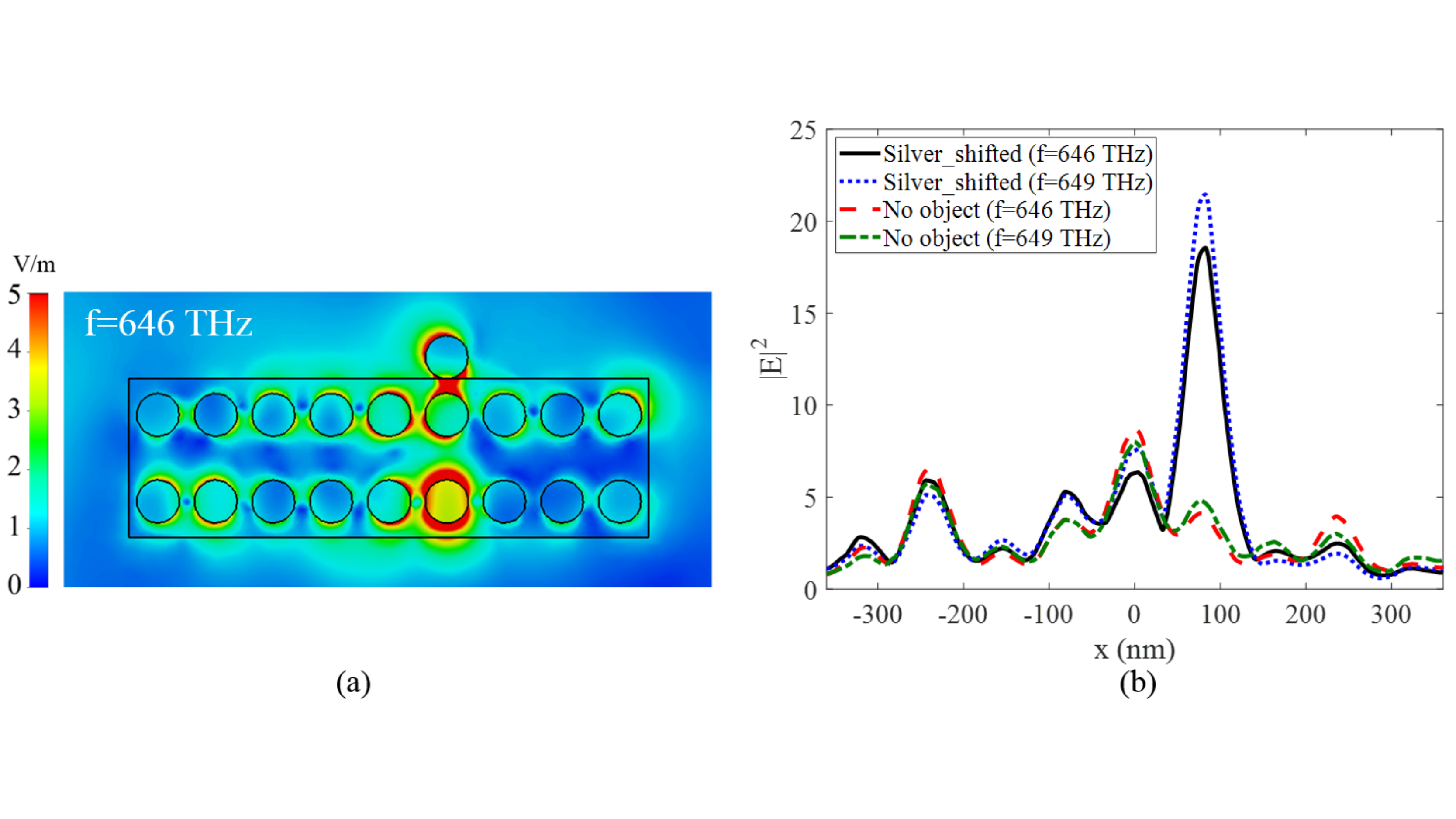}
 \caption{(a) Color map of the electric field amplitude in the central vertical plane ($y=Na/2$) of the superlens in presence of a silver sphere with
 $d_1=d$ on top of the superlens.
 (b) Electric intensity profile along the intersection line of the planes $y=Na/2$ and $z=-h-h_1-h_2$.}
 \label{fig8}
\end{figure*}

let us discuss the novelty brought by these investigations compared to our previous studies \cite{Alitalo,OL}. First, in the previous works the nanoimaging
was numerically demonstrated for the vertically polarized sources, perhaps forming complex-shape clusters as in \cite{OL}. These sources
offered the domination of the vertical polarization in the superlens and the idea of the superlens was to exploit the collective resonance of vertically polarized nanoparticles .

In the present work, we reject the idea to consider the tangential polarization as a parasitic one (adopted in works \cite{Mas,Mar,Sim1,Syd,Syd1})
because instead of optimizing a single metasurface engineering its dispersion curve maximally close to Fig.~\ref{fig2}(a),
we optimize the dual metasurface and aim to obtain the bandgap between the frequency regions of SPPs without taking care on the shape of the dispersion curves in these regions.

For our superlens it does not matter what to image -- radiating sources or scattering objects, nanospheres or nanorods.
We will see below that the same nanoimaging parameters keep for scatterers of different materials (Ag, PEC and dielectric of permittivity $\va=3$).

Next, in our present simulations the incident wave uniformly illuminates the whole superlens up to its edges and the modes related to the dimensional effects may be excited.
In the simulations of \cite{Alitalo,OL} where the sources were located in the central area of the superlens
the sufficiently distant edges were practically not excited by the field sources. Moreover, now we take into account the host medium interfaces which
impact was absent in our previous works. It is important to show that these factors do not worsen the imaging at all.

What worsens the imaging is the parasitic local field enhancement in the image area at the superlens frequencies. In the simplistic model, the
plasmonic hot spots extended into this area only at the frequencies of the SPPs. So, our present simulations drastically change the previous insight of a dual-metasurface plasmonic superlens.

\begin{figure*}[t]
\centering
\includegraphics[width=0.9\textwidth]{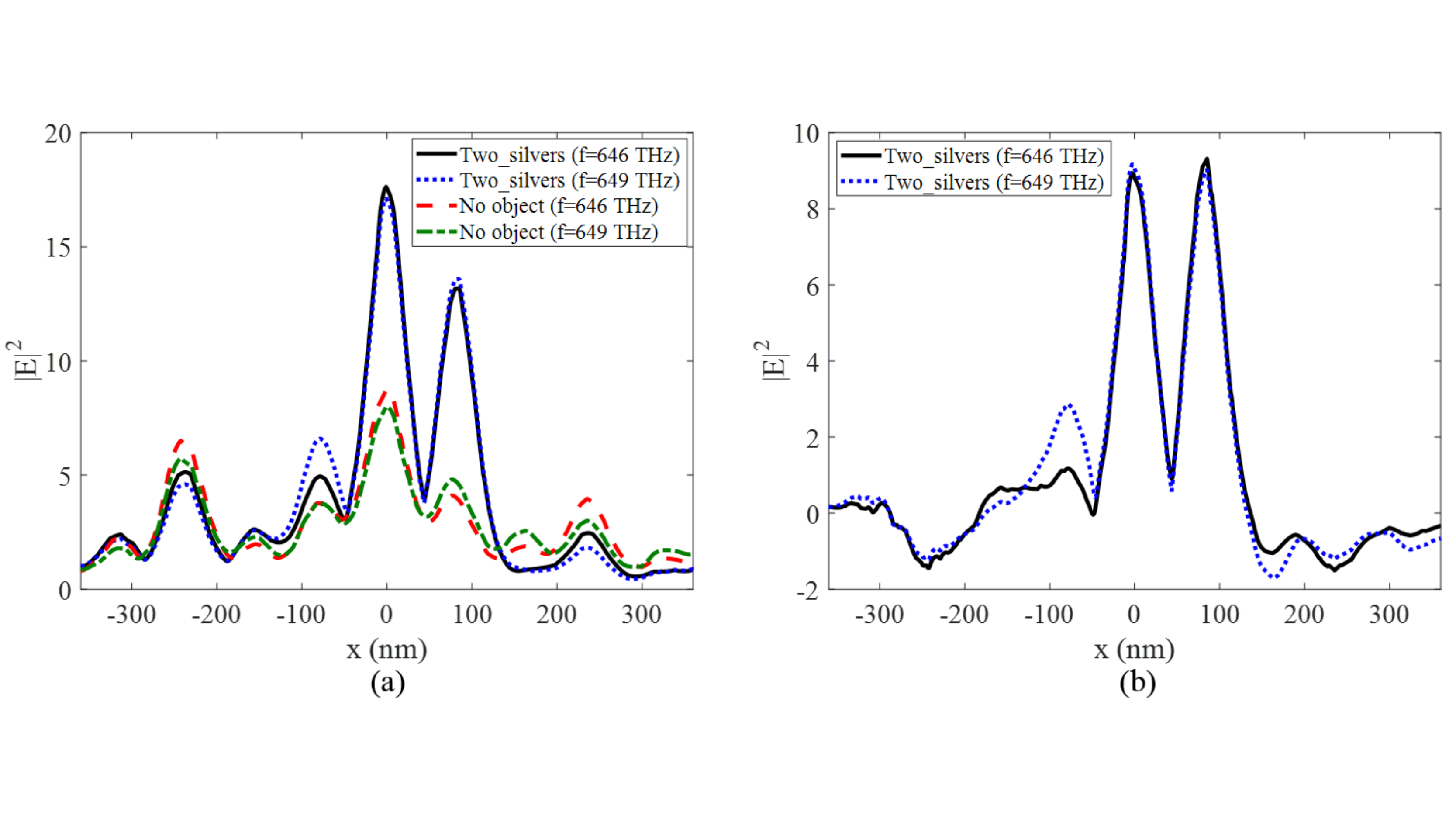}
\caption{(a) Electric intensity profile along the intersection line of the planes $y=Na/2$ and $z=-h-h_1-h_2$ in presence of two silver spheres with $d_1=d$.
(b) The same intensity profile from which the parasitic signal is subtracted.}
\label{fig10}
\end{figure*}

The variety of the materials and shapes of the subwavelength objects, the variation of the incidence angle and the variation of $N$
allowed us to establish the general features of the imaging that holds for our superlens (see above) in the range 646-649 THz:
\begin{itemize}
\item
The nanoimaging holds only for TM-polarized incident waves and demands a sufficient incidence angle $\theta$.
\item
The plane in which the image field forms the highest contrast with the parasitic background field
is slightly distanced from the bottom surface of the superlens, but it may be approximately considered
as the bottom surface of the superlens if the objects located on top of the superlens are small enough.
\item
The contrast of the imaging and the background fields does not depend on $N$ if $N$ is sufficiently large and
rather weakly depends on the incidence angle $\theta$. However, it noticeably depends on the material of the nanoobjects.
\item
The ratio between the transversal sizes of the image and the object (image magnification) as well as the spatial resolution of two subwavelength objects
in the horizontal plane do not depend even on the shape and the material of the objects. For a given incidence wave within the operation band
both image magnification and resolution are operation parameters of the superlens.
\item
There is an optimal incidence plane (bi-sectorial with respect to the Cartesian axes) and the optimal incidence angle ($\theta=\pi/4$) for which the
the object shape in the plane $(x-y)$ is not distorted and the image magnification is nearly equal unity.
\item
The interfaces of the finite plate hosting the dual metasurfaces do not worsen the imaging.
\end{itemize}

\begin{figure}[t]
\centering
\includegraphics[width=0.45\textwidth]{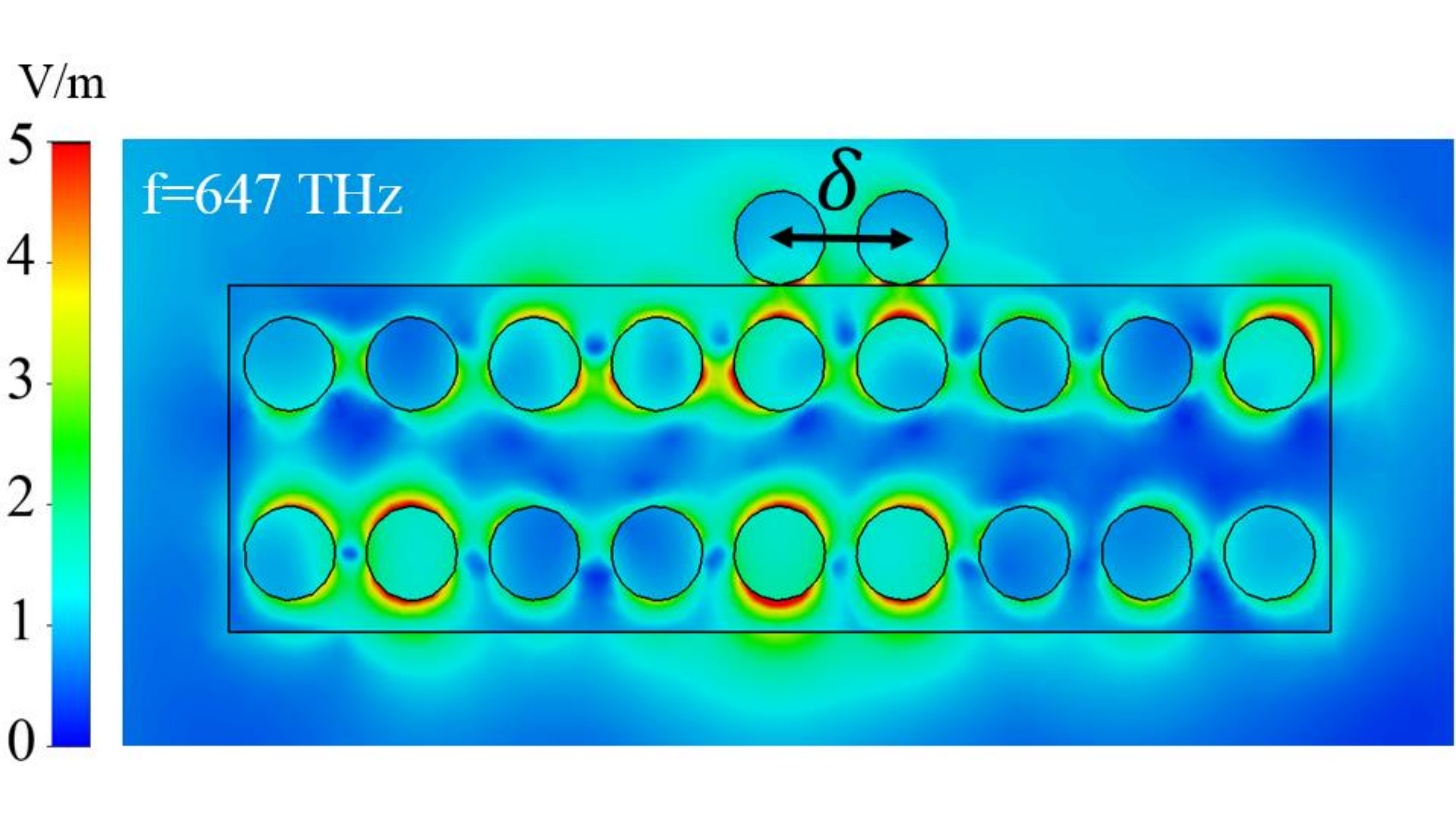}
\caption{Color map of the electric field amplitude in the central vertical plane of the superlens in presence of two dielectric
spheres with $d_1=d$ and $\va_{r}=3$.}
\label{fig12}
\end{figure}

In Fig.~\ref{fig8} we present the color map of the electric field in the vertical cross section of the superlens at 649 THz and the field intensity profile
along $x$ in the image plane at two frequencies (646 and 649 THz). These plots correspond to the nanoimaging of one Ag nanosphere of size $d_1=60$ nm.
For comparison we shown in Fig.~\ref{fig8}(b) the intensity profiles at the same frequencies in absence of the object. We can see that for the objects
which are themselves resonant in the superlens operation band the local field enhancement at the superlens frequencies
is not an issue. The image field is much larger than the parasitic one.

\begin{figure*}[t]
\centering
\includegraphics[width=0.9\textwidth]{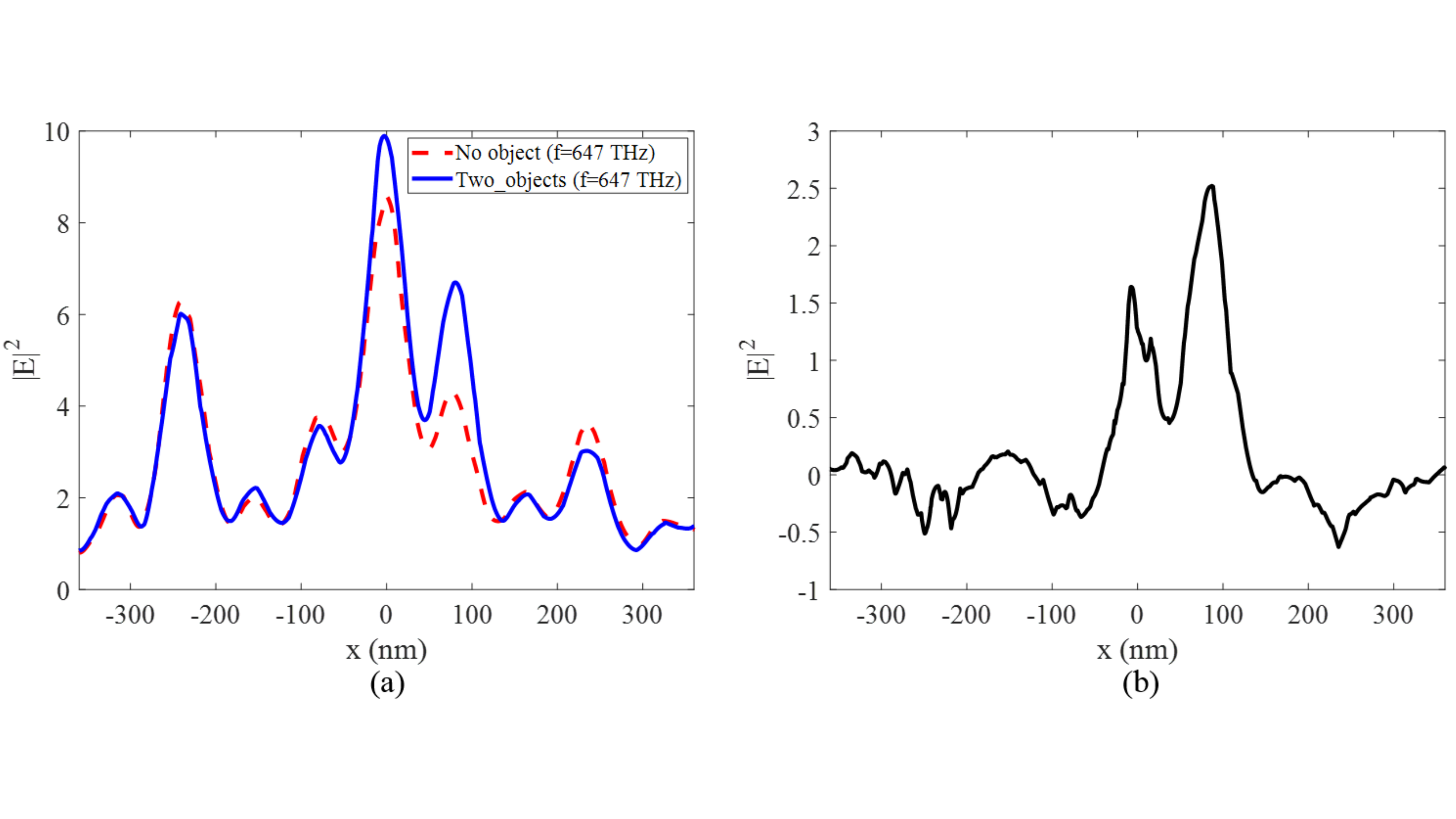}
\caption{(a) Electric intensity profile along $x$ in the image plane in presence of two dielectric spheres.
(b) The same intensity profile from which the parasitic signal is subtracted.}
\label{fig13}
\end{figure*}

However, even for silver nanoobjects, if we want to resolve them, the subtraction of the parasitic signal would be useful. In Fig.~\ref{fig10}(a) we show the
field intensity profiles along $x$ in the image plane in the case of two Ag objects separated by the distance $80$ nm. This is practically the minimal distance
$\delta$ still corresponding to the Rayleigh criterion of spatial resolution. When the distance is reduced to $75$ nm the overlapping of two images is overcritical.
Thus, we can put $\delta=a$, that is confirmed by the analysis of the color maps in the vertical cross sections and in the image plane.
However, the peaks corresponding to two spheres in Fig.~\ref{fig10}(a) are of different height -- the background field, though small, is not negligible and
the objects are located not symmetrically with respect to the superlens center. When we subtract the parasitic intensity and depict the differential
intensity profile in Fig.~\ref{fig10}(b) both peaks in the superlens band are almost identical and this differential image is not distorted.

For nanoimaging of the non-resonant objects, subtracting the parasitic signal is necessary. In Fig.~\ref{fig12} we see that the
color map of the electric field amplitude in the central vertical plane of the superlens in presence of two dielectric objects.
The spots corresponding to the objects in the image plane visually overlap and it is not evident that these objects are resolved.
They seem to be resolved in the intensity profile depicted in Fig.~\ref{fig13}(a), but it is not so.
The Rayleigh criterion demands that the intensity in the middle between two peaks (centers of two spots imaging the two point sources)
would be at least one half of the intensity in both peaks. In the right peak the normalized intensity is equal $6.7$ and in the middle it is equal $3.8$
i.e. the Rayleigh criterion is not respected. After subtracting the parasitic signal we see in the differential intensity profile depicted in
Fig.~\ref{fig13}(b) that the Rayleigh resolution holds for the differential intensity profile.
The frequency in these plots is that of the best superlens operation (647 THz), and the resolution
highly exceeds the Rayleigh criterion. At 646 and 649 THz the Rayleigh resolution still holds for $\delta=a$.
Our superlens is an aberration-free near-field imaging device with the spatial resolution $\delta\approx a\approx \lambda/6$, where $\lambda$ is the central wavelength of the operation band.
\begin{table}[h!]
	\caption{{Comparison of superlenses.}}
	\centering
	\begin{tabular}{||c||c|c|c||}
		\hline\hline
		Superlens & work \cite{Alitalo} & work \cite{OL} & This paper \\ [0.5ex]
		\hline
		Performance & $\delta = 0.34 \lambda$ & $\delta = 0.31 \lambda$ & $\delta = 0.16 \lambda$ \\
		parameters & $D = 0.55 \lambda$ & $D = 0.63 \lambda$ & $D = 0.56 \lambda$ \\ [0.5ex]
		Design & $d = 56$ & $d = 60$ & $d = 60$ \\
		parameters &$d_z = d$  & $d_z = 30$ &  $d_z = d$\\
		(nm)& $a = 65$ & $a = 65$ & $a = 80$ \\
		& $h = 130$ & $h = 130$ & $h = 120$ \\ [1ex]
		\hline\hline
	\end{tabular}
\end{table}

To sum this part up, we have numerically proved (for the first time) that the plasmonic superlens
images with subwavelength resolution not only vertical sources but also scattering objects and
(in contrast to the previous claims of \cite{Mas,Mar,Alitalo,OL,Sim1,Syd,Syd1}) its
operation is related not with the vertical polarization of the eigenmodes but with the bandgap between two SPPs having the arbitrary tilted polarizations.
This new insight allowed us to properly optimize the superlens.
The design parameters found from analytical calculations were mainly confirmed by full-wave simulations.
The correct optimization offered us the better output parameters compared to
\cite{Alitalo,OL}. Namely, for $D=260$ nm or $0.56\lambda$ (nearly the same distance from the source to the image plane as in \cite{Alitalo,OL}) we have obtained
$\delta\approx \lambda/6$ instead of $\delta\approx \lambda/3$ i.e. twice higher spatial resolution. {A comparison between these new results and results of works \cite{Alitalo,OL} is presented in Table I.}

\begin{figure}[t]
	\centering
	\includegraphics[width=0.49\textwidth]{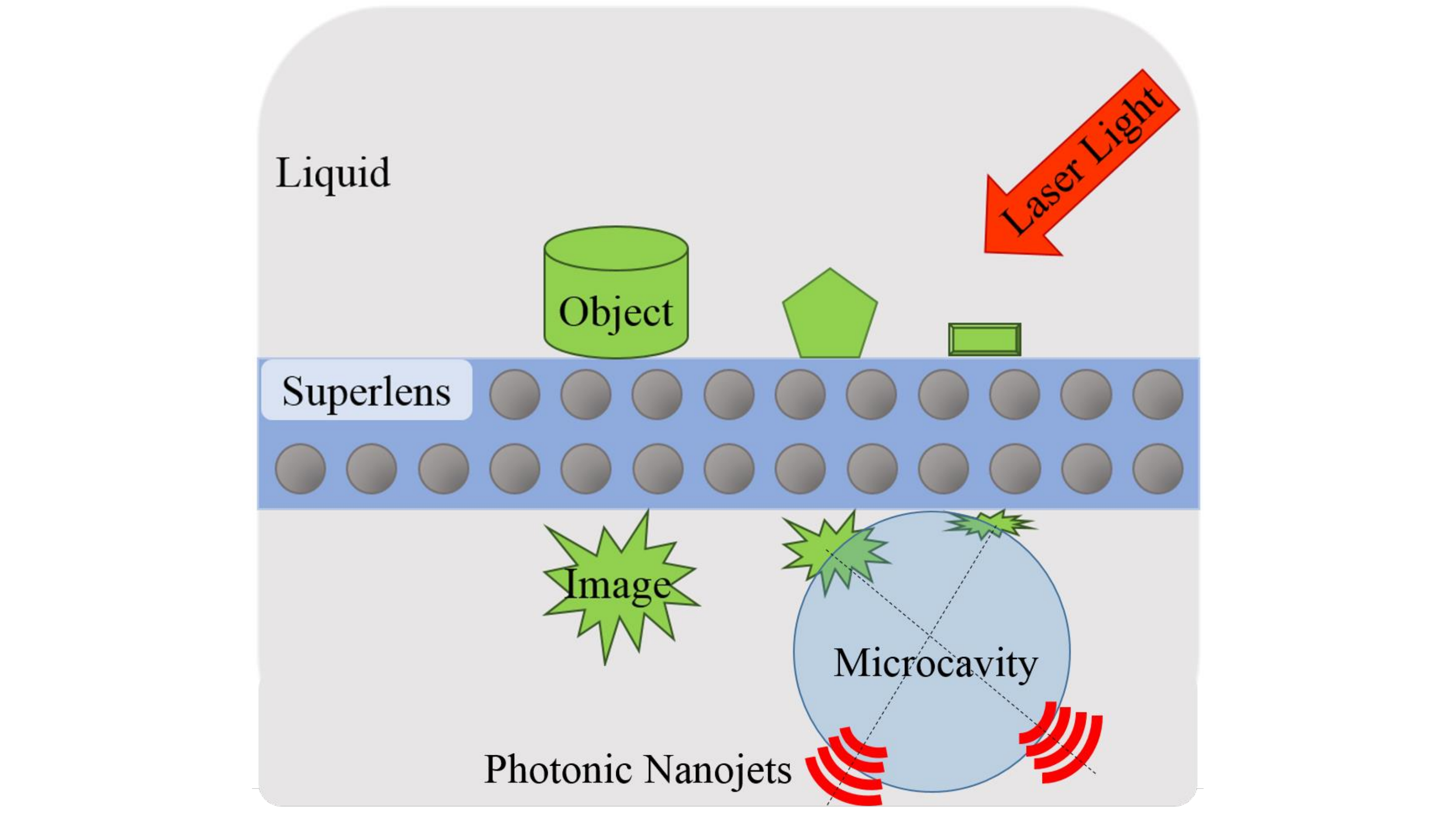}
	\caption{Conceptual scheme of the far-field imaging assisted by our superlens.}
	\label{fig3}
\end{figure}

\section{Conclusion}

In this work we have completely clarified the physics underlying the operation of a dual-metasurface plasmonic superlens.
For the first time, the nanoimaging of a dual-metasurface superlens was shown using full-wave simulations instead of a point-dipole model,
for scattering objects illuminated by a plane wave instead of radiating sources, and for a realistic design of the dual metasurface encapsulated into a finite glass plate
instead of a uniform ambient. {A scientific novelty of this work is the replacement of a simplistic
analytical model developed earlier. The simplistic model is suitable when only one -- vertical
polarization -- exists in the structure, as it holds for microwave superlens of wire loops \cite{Syd1}. It is qualitatively adequate, if this polarization is
resonantly excited at the operation frequency and therefore dominates over the horizontal one, as it holds for a superlens of plasmonic nanotablets studied in \cite{OL}.
The simplistic model targets the maximally flat dispersion curve in the range of surface waves. However, as we have seen this target does not grant the true optimization
because the dispersion in the dual metasurface turns not flat and has little common features with that of the single metasurface.
The true optimization is the search of the maximally wide bandgap for the SPPs in the dual metasurface.
Aiming this target we have drastically improved the operation characteristics of the superlens.}

In Appendix we suggest a novel application for the superlens operating in the visible range, that allows the magnification of the subwavelength image
without mechanic or any other disturbing the object area. Finally, let us mention that we have tried to design a dual-metasurface superlens using golden nanoparticles.
However, an insufficient figure of merit of Au in the plasmon frequency range did not allow us to engineer the necessary bangap between two dispersion curves.

\begin{figure*}[t]
	\centering
	\includegraphics[width=0.75\textwidth]{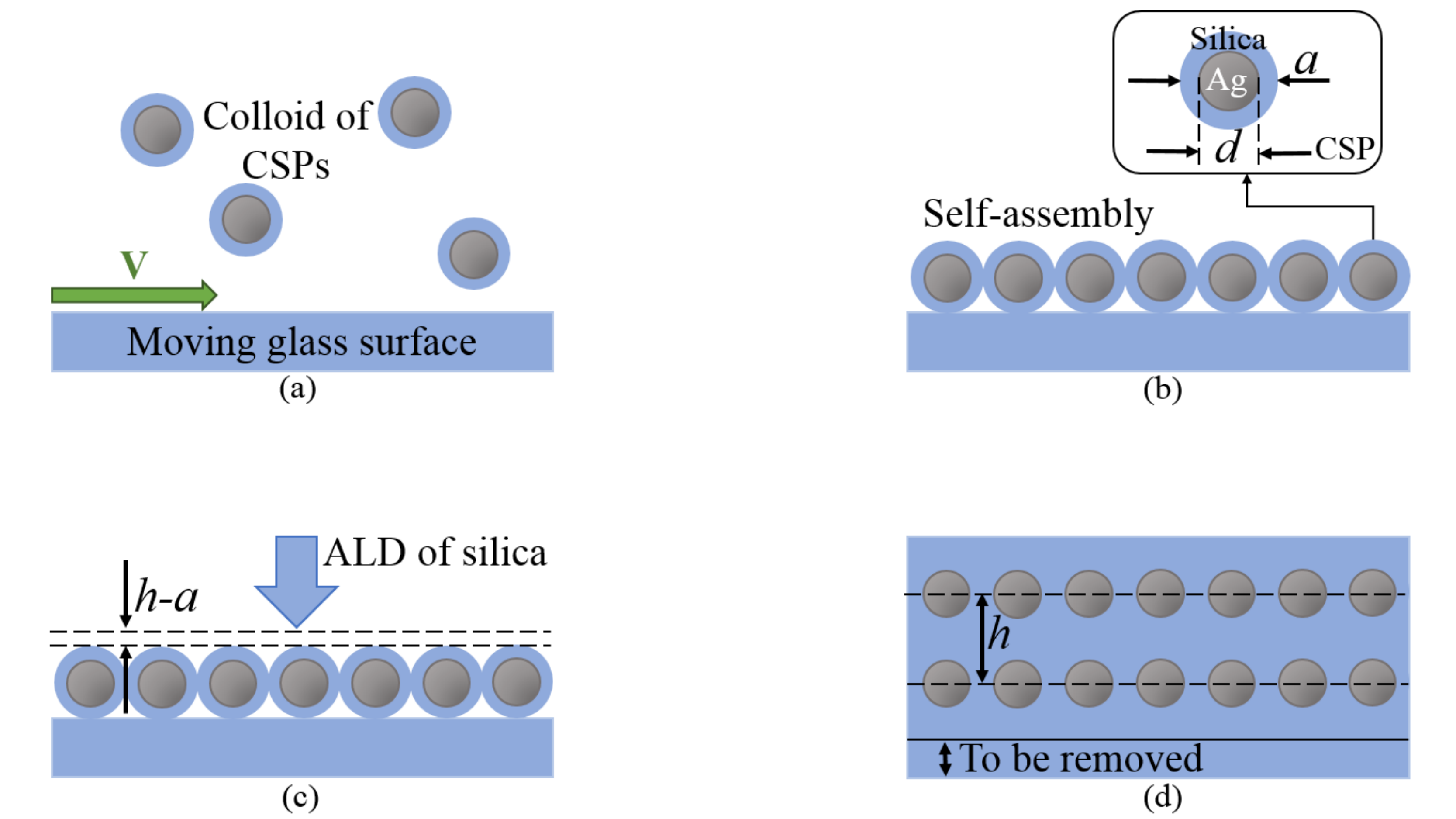}
	\caption{Suggested fabrication technique for fabricating the superlens and atomic layer deposition:
		(a) moving a glass plate with a proper speed $v$ we enable the self-assembly of core-shell plasmonic nanospheres from a colloidal suspension;
		(b) the result of this self-assembly is a densely-packed monolayer; (c) the monolayer of nanospheres is impinged by silica in the ALD machine (silica ideally fills
		all nanogaps and allows to define the needed thickness $h-a$ on top of the monolayer), (d) preparing in the same
		way the second layer of Ag nanospheres embedded into glass we finally remove the unnecessary bottom part of the glass plate
		and obtain a dual-metasurface superlens.}
	\label{fig4}
\end{figure*}

\section*{Acknowledgement}

This work was supported by the European Association of National Metrological Institutes, project number 17FUN01 (Light-matter interplay for optical metrology beyond the classical spatial resolution limits).

\section*{Appendix}

Our target application is a label-free imaging of submicron biological objects which can move in a liquid medium.
The goal is to remotely see and resolve these objects when they approach to a dielectric surface. We need the fast imaging in real time
and cannot introduce any disturbance into the area of our objects. If the object image is obtained on the back side of a
superlens which simultaneously serves a membrane separating the area of the object from that of the image
we can see and resolve the magnified subwavelength images of the objects. The conceptual scheme of this imaging technique is presented in Fig.~\ref{fig3}.
The image of the object is a hot spot or, alternatively, the package of evanescent waves. It excites the microcavity in the same way as a nearly-located scatterer
and a notch do. Both diameter $d$ and refractive index $n$ should be chosen so that the superlens operation frequency would be that of the WGR.

Further, we suggest a novel fabrication technique for our dual-metasurface superlens allowing its affordable implementation in a rather large area using the self-assembly of plasmonic
nanoparticle from a colloid and the atomic layer deposition of silica. The suggested fabrication technique is illustrated by Fig.~\ref{fig4}. We have already tested this approach preparing the light-trapping structures for thin-film solar cells in works \cite{MO1,MO2}. Colloidal suspensions of core-shell nanopsheres with Ag core and silica shell shown in the inset of Fig.~\ref{fig4}
are available with broad range of the core diameters $d$ and the shell thicknesses $(a-d)/2$. As it is explained previously, we need $d=50-70$ nm and $(a-d)/2=10-15$ nm. Moving a glass plate with a proper speed $v$ {(to be found experimentally)} inside the colloid we will enable the self-assembly of the densely packed monolayer of silica shells on glass. It results in the formation of a densely packed monolayer of nanospheres. After drying the structure we will perform the ALD of silica on top of the structure. The deposited molecules will ideally fill all the nanogaps (see in \cite{MO2}) and we will obtain a flat glass layer of thickness $h-a$ above the nanospheres. Then we will put this glass plate with incorporated Ag nanospheres into the same colloidal suspension and repeat the same steps until we obtain the dual-metasurface structure depicted in Fig.~\ref{fig4}(d). The extra amount of glass on the bottom of the plate can be easily removed by chemical or ion-beam etching.
This fabrication technique is more promising compared to one corresponding to a hyperbolic-medium superlens and makes our superlens favorable.

\end{document}